\documentclass[aps,floatfix,twocolumn]{revtex4}

\usepackage{graphicx}
\graphicspath{{figures/}}

\usepackage{amsmath,amssymb}
\usepackage{units,xspace}

\renewcommand{\vec}[1]{\boldsymbol{#1}}
\newcommand{\vecr}{\vec{r}}
\newcommand{\abs}[1]{\left\vert #1 \right\vert}
\newcommand{\avg}[1]{\left< #1 \right>}
\newcommand{\micron}{\ensuremath{\unit{\mu m}}\xspace}

\begin{document}

\title{Holographic optical trapping}

\author{David G. Grier}

\author{Yael Roichman}

\affiliation{Department of Physics and Center for Soft Matter Research,
New York University, 4 Washington Place, New York, NY 10003}

\begin{abstract}
Holographic optical tweezers use computer-generated holograms to
create arbitrary three-dimensional configurations of single-beam
optical traps useful for capturing, moving and transforming 
mesoscopic objects.  Through a combination of
beam-splitting, mode forming, and adaptive wavefront correction,
holographic traps can exert precisely specified and characterized
forces and torques on objects ranging in size from a few nanometers
to hundreds of micrometers.  With nanometer-scale spatial resolution and
real-time reconfigurability, holographic optical traps offer
extraordinary access to the microscopic world and already have found
applications in fundamental research and industrial applications.

Key words: Optical trapping, computational holography, digital video
microscopy
\end{abstract}

\maketitle

\section{Introduction}
Two decades after their invention, single-beam optical gradient force traps, commonly known as optical
tweezers, have become indispensable tools for research
\cite{ashkin86,ashkin00}.
Formed by bringing an intense beam of light to a diffraction-limited
focus, an optical tweezer can capture an object ranging in
size from a few nanometers to several micrometers and hold it stably
in three dimensions against gravity, random thermal forces, and other
external influences.
This article addresses a generalization of the optical tweezer
technique that uses computer-generated holograms (CGH) to create
hundreds of simultaneous optical tweezers in arbitrary 
three-dimensional configurations, each with individually specified 
trapping characteristics.
Introduced in 1997, holographic optical traps 
\cite{dufresne98,reicherter99,liesener00,dufresne01a,curtis02,polin05}
have found applications in research and engineering ranging from
fundamental studies of the mechanisms of phase transitions to the
manufacture of wavelength-scale devices \cite{grier03}.

A single optical tweezer works by minimizing the electromagnetic
energy stored in the fields scattered and absorbed
by an illuminated object \cite{maianeto00}.
Generally, this results in a small object being localized near the
focus of a strongly converging laser beam.
Heuristically, and semiquantitatively for sub-wavelength-scale Rayleigh objects,
the attractive force may be understood as arising from a dipole moment
induced in the particle by the light's fields.
The induced dipole 
is drawn up gradients of the field toward the focus,
where the light is brightest.
Because the induced dipole moment typically is proportional to the
field 
and the force on the dipole is 
proportional to the local field gradient, the
overall trapping force is proportional to gradients in the intensity.
This insight is exploited in the next section to simplify the creation of holographic trapping
arrays.

Radiation pressure due to
absorption and backscattering
competes with the attractive gradient force and tends to blow
particles downstream.
Stable three-dimensional trapping in a single beam of light is
possible
only if the axial intensity gradients are large enough to overcome
radiation pressure.
This is one reason that optical tweezers generally are created with
high-numerical-aperture
lenses, such as microscope objectives, that are capable of bringing
a beam of light to an exceptionally tight focus.
Geometric aberrations degrade an optical tweezer's 
performance by reducing the focal spot's intensity gradients.
Microscope objectives' well corrected aberrations also recommend them
for this application.

A single collimated beam that fills an infinity-corrected
objective's input pupil comes to a focus and forms a trap 
in the lens' focal plane
at a position dictated by the beam's angle of incidence.
Any object trapped in the tweezer therefore can be imaged
conveniently with the same lens, provided that the imaging
illumination can be separated from the trap-forming laser,
for instance with a dichroic mirror.
A diverging beam filling the lens' input pupil forms a trap downstream
of the focal plane, and a converging beam forms a trap upstream.
Controlling the input beam's degree of collimation and angle of incidence therefore
provides a mechanism for positioning an optical tweezer in
three dimensions.

\section{Holographic optical trapping}

\begin{figure}[htbp]
  \centering
  \includegraphics[width=3in]{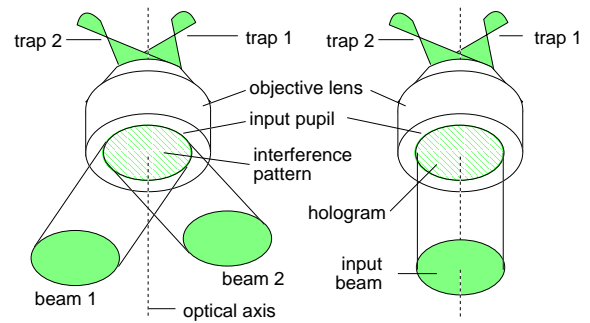}
  \caption{Two beams of light focus to two optical tweezers,
  and also form an interference pattern at the lens' input pupil.  The
  same traps can be created from a single input beam by placing an 
  equivalent hologram in the input pupil.
  }
  \label{fig:lens}
\end{figure}
Multiple beams of light all passing through the objective's input
pupil with their own angle of incidence and degree of collimation
create a configuration of optical traps, as shown in Fig.~\ref{fig:lens}.
If these beams are mutually coherent, they form an interference
pattern in the input pupil, with fields of the form
\begin{equation}
  \label{eq:fields}
  \psi(\vec{\rho}) = u(\vec{\rho}) \, \exp\left( - i \, \varphi(\vec{\rho}) \right),
\end{equation}
at point $\vec{\rho}$.
Were the same pattern of amplitude modulations, $u(\vec{\rho})$,
and phase modulations, $\varphi(\vec{\rho})$,
imposed on the wavefronts of a single incident beam as it passed
through the input pupil, the
modified beam also would create the same trapping pattern.
This is the principle behind holographic optical trapping.

Creating multiple optical traps does not require a fully complex
hologram.
Because optical trapping relies only on gradients in the intensity,
and not on the phase, even quite complicated three-dimensional
configurations
of optical traps can be specified with just $\varphi(\vec{\rho})$,
leaving the amplitude profile $u(\vec{\rho}) = u_0(\vec{\rho})$ of the
input beam unchanged.
The phase-only diffractive optical element (DOE) encoding a particular
pattern of traps is an example of a class of holograms known as
kinoforms.
The trick, then, is to compute the kinoform that projects a
particular pattern of traps.

Several algorithms have been proposed for seeking holograms
that most accurately and most rapidly approximate desired trapping patterns.
The fastest is to compute the phase associated with a linear superposition
of the desired beams, and to simply discard the associated amplitude
variations \cite{liesener00}.  Such straightforward superposition is surprisingly effective,
particularly if the beams are chosen to have random relative phases.
The resulting trapping pattern tends to be marred, however,
by large numbers
of ``ghost'' traps at symmetry-dictated positions, and also by large variations
in the traps' intensities from their design values.
For many applications, however, the resulting performance is more than adequate, and
the ease of computation facilitates real-time interactive control.

Superposition also provides an outstanding starting point for refinement algorithms.
Iterative refinement schemes based on the Gerchberg-Saxton and adaptive additive
algorithms \cite{soifer97} improve all aspects of the holograms' performance \cite{dufresne01a,sinclair04}, 
although at substantial
computational cost, particularly for three-dimensional trapping patterns.
A modified adaptive-additive algorithm that calculates fields only at the
traps' locations \cite{curtis02} is far more efficient, but also less effective at suppressing
ghost traps.  More recently, direct search algorithms have been shown to yield
substantially more accurate DOE estimates \cite{polin05} and also can be far
more efficient
if started from the randomly-phased superposition \cite{polin05} rather than from
a random phase field \cite{sinclair04}.

\begin{figure}[htbp]
  \centering
  \includegraphics[width=3in]{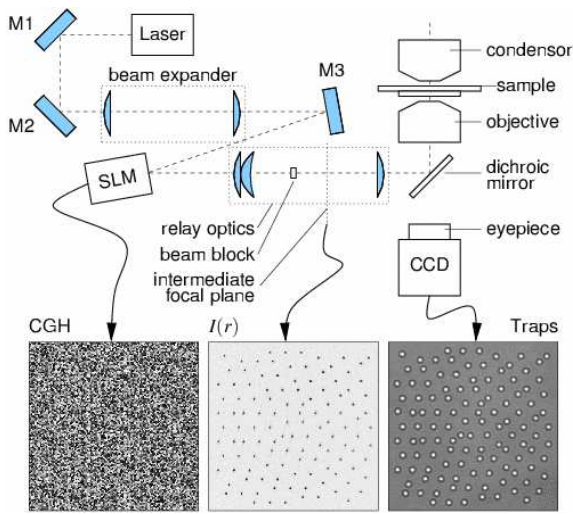}
  \caption{Schematic implementation of holographic optical traps.  An
    expanded laser beam is reflected by a liquid crystal spatial light
    modulation, which imprints a computer-generated hologram onto its
    wavefronts.  The $200 \times 200$ pixel region of a CGH shown
    encodes a pattern of 119 optical tweezers in a quasiperiodic
    arrangement.  The phase hologram is relayed to the input pupil
    of an objective lens that focuses it into holographic optical
    traps, shown here trapping 1.5~\micron diameter colloidal spheres
    in water.}
  \label{fig:optics}
\end{figure}

The field due to an array of $M$ discrete point-like traps located at $\{\vecr_m\}$ 
can be approximated by
\begin{align}
  \label{eq:sumoffields}
  \psi(\vecr) & = \sum_{m=1}^M \psi_m \, \delta(\vecr - \vecr_m) \text{, \quad with}\\
  \psi_m & = \alpha_m \, \exp(-i \phi_m),
\end{align}
where $\alpha_m$ is the relative amplitude of the $m$-th trap,
normalized by $\sum_{m=1}^M \abs{\alpha_m}^2 = 1$.
The relative phases, $\phi_m$, generally are assigned randomly, but also may
be specified for particular applications.
In most practical implementations, such as that depicted in Fig. \ref{fig:optics},
the DOE, $\varphi(\vec{\rho})$, encoding the traps also is discretized into
an array of $N$ phase pixels $\varphi_j$ located at $\vec{\rho}_j$.
Consequently, the complex field at each trap can be described 
by a nonlinear transformation of the input phase
\begin{equation}
  \label{eq:transfer}
  \psi_m = \sum_{j=1}^N T_{m,j} \, u_j \, \exp(i \varphi_j),
\end{equation}
where the transfer matrix $T_{m,j}$ describes the coherent propagation
of light from pixel $j$ on the DOE to trap $m$ in the focal plane,
given the input beam's amplitude profile,
$u_j = u_0(\vec{\rho}_j)$.
In our implementation, the amplitude profile is approximated by the Heaviside
step function $u_0(\vec{\rho}) = \Theta(\rho - R)$, where $R$ is the radius of the optical
train's aperture.

The transfer matrix for a two-dimensional configuration of 
conventional optical tweezers is given in scalar diffraction theory by
\cite{goodman96,polin05}
\begin{equation}
  \label{eq:fresnel}
  T^{(0)}_{m,j} = \frac{1}{\lambda f} \, 
  \exp\left(-i \frac{2\pi \vecr_m \cdot \vec{\rho}_j}{\lambda f} \right).
\end{equation}
More generally, the transfer matrix can take the form
\begin{equation}
  \label{eq:transfermatrix}
  T_{m,j} = \prod_{k=0}^{K_m} T^{(k)}_{m,j},
\end{equation}
where the additional $K_m$ contributions, $T^{(k)}_{m,j}$, describe wavefront-shaping
operations specific to the $m$-th trap.  For example, if the DOE
displaces the $m$-th trap by a distance $z_m$ along the
optical axis, then
\begin{equation}
  \label{eq:displace}
  T^z_{m,j} = \exp\left(i \, \frac{2\pi \rho_j^2 z_m}{f^2}\right)
\end{equation}
returns its image to the focal plane for analysis \cite{curtis02,polin05}.
More dramatic transformations implemented with Eq.~(\ref{eq:transfermatrix})
will be described in Sec.~\ref{sec:multimode}.


Direct search refinement starts from an estimate
$\varphi_j$ for the DOE, and the associated fields $\psi_m$
calculated with Eq.~(\ref{eq:transfer}).
If the DOE exactly encoded the desired trapping pattern, then
the calculated amplitudes $\abs{\psi_m}$ would agree with the
design values, $\alpha_m$.  
The algorithm seeks to minimize actual discrepancies between $\abs{\psi_m}$
and $\alpha_m$.
Following Meister and Winfield \cite{meister02}, we adopt the error function
\begin{equation}
  \label{eq:error}
  E = - \avg{\abs{\psi_m}^2} + \gamma \,
  \sqrt{\avg{\left(\abs{\psi_m}^2 - \alpha_m^2 \,\frac{\avg{\abs{\psi_m}^2 \alpha_m^2}}{\avg{\alpha_m^4}} \right)^2}},
\end{equation}
where the weighting factor
$\gamma$ sets the relative importance attached to diffraction efficiency ($\gamma = 0$) and fidelity to
design ($\gamma > 0$).
Improvements are sought by selecting pixels at random, changing
their phase values, recomputing the fields, and retaining only
those proposed changes that improve the performance.
Because the relationship between $\abs{\psi_m}$ and $\varphi_j$
is inherently nonlinear, the search proceeds sequentially,
and the process continues until $E$ is reduced to an acceptable
level.

In practice, convergence starting from a randomly-phased superposition 
typically is achieved with a single pass through the array, for a total of $MN$ operations.
This is comparable in computational cost to the initial superposition and so roughly
doubles the total cost of the computation.
The benefits of the refinement step can be substantial, as a practical example
illustrates.

We project holographic optical traps with the system shown
schematically in Fig.~\ref{fig:optics}.
Light from a frequency-doubled Nd:YVO$_4$ laser (Coherent Verdi) is expanded to fill the face
of a reflective liquid crystal spatial light modulator (SLM) (Hamamatsu X8267-16 PPM), which can impose a
phase shift between 0 and $2 \pi$ radians at each pixel in a $768 \times 768$ array.
The phase-modified beam is relayed to the input pupil of a 100$\times$, NA 1.4, SPlan Apo oil
immersion objective mounted in a Nikon TE-2000U inverted optical microscope, which focuses
the light into optical traps.
Because the SLM's face is in a plane conjugate to the objective's input pupil, the effect
is the same as if the DOE were placed in the input pupil, as in Fig.~\ref{fig:lens}.
The benefit of this arrangement is that the trapped sample can be imaged onto a CCD camera
using the microscope's standard optical train, with the imaging illumination passing through
the dichroic mirror used to direct the trap-forming laser.

In practice, not all of the input beam is diffracted by the SLM, and the undiffracted portion
ordinarily would form a bright trap right in the middle of the field of view.
To counter this, we adjust the beam expander so that the SLM is illuminated with a slightly
converging beam.  
Projecting optical traps into the microscope's focal plane therefore requires the traps
to be displaced along the optical axis with the computed DOE.  The undiffracted beam
therefore focuses into a different plane within the relay optics than the intended traps,
and so can be blocked with a spatial filter without disrupting the traps.
Displacing the trapping plane has the additional benefit of projecting most residual
ghost traps out of the sample volume.

The result can be seen in the typical images in Fig.~\ref{fig:optics}.
Here, an eight-bit CGH imprinted on the input beam by the SLM creates the
pattern of focal spots in the intermediate focal plane, which is shown trapping
colloidal spheres dispersed in water.
This particular quasiperiodic arrangement of 119 optical traps is particularly
challenging because it lacks reflection symmetry about the optical axis.
As a result, a typical DOE computed by superposition alone suffers from more than
50 percent root-mean-squared (RMS) relative deviations from design amplitudes.
Imaging photometry and measurements of the traps' potential energy wells by particle
tracking \cite{polin05,crocker96} confirm that the DOE refined by direct search 
is uniform to within 5 percent, a factor of ten improvement.

\begin{figure}[htbp]
  \centering
  \includegraphics[width=3in]{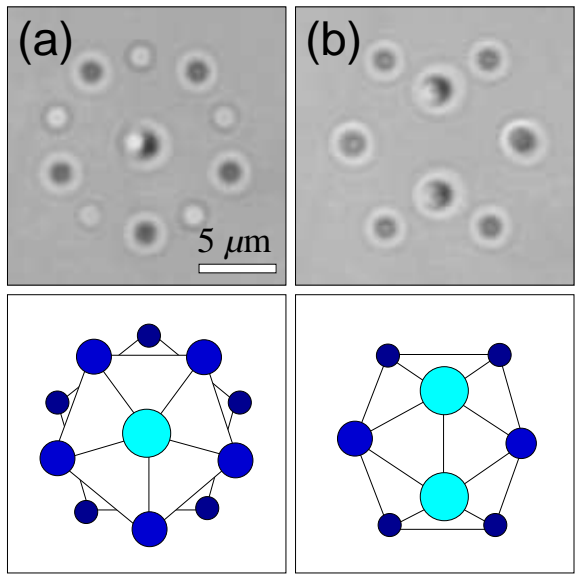}
  \caption{Two views of a rotating icosahedron of
    colloidal spheres created with dynamic holographic optical tweezers.}
  \label{fig:icosahedron}
\end{figure}
Our implementation of dynamic holographic optical trapping
permits full three-dimensional manipulation over a $100 \times 100 \times 40~\unit{\micron^3}$
volume.  Micrometer-scale colloidal spheres are readily stacked five or more deep along the
axial direction, with three-dimensional quasicrystals consisting of hundreds of spheres
having recently been demonstrated \cite{roichman05}.
The arrays' fidelity to design intensities, and the DOE's overall efficiency fall off as
the arrays become increasingly complicated.
How design complexity affects implementational
efficacy has yet to be
worked out.

Demonstrations of three-dimensional control \cite{liesener00,sinclair04} 
such as the rotating icosahedron in
Fig.~\ref{fig:icosahedron},
reveal that objects can be organized into vertical stacks
along the optical axis.
Three-dimensional assemblies consisting of hundreds of spheres
in asymmetric configurations up to nine layers deep recently
have been demonstrated \cite{roichman05}.
Still larger areas and depths can be accessed, at least in principle,
by creating time-shared three-dimensional trapping patterns
\cite{melville03}.
The resulting structures can be made permanent, for example by gelling the
suspending fluid \cite{korda02,jordan04}.

\section{Static optical landscapes: Transport and fractionation}

Dynamic optical trapping arrays have immediate applications for manipulating
microscopic objects such as biological cells and organizing them into
useful and interesting configurations.
The speed with which patterns consisting of hundreds of simultaneous traps can be
animated with conventional computing hardware lends itself to interactive
real-time manipulation, as indeed is the case for the commercial implementation
of this technology (BioRyx 200 system, Arryx, Inc.).
Even static optical trapping arrays, however, have surprising and
practical applications.

Each optical trap in an array acts as a three-dimensional potential
energy well for a small object.  An entire array, therefore, may be
viewed as an extended potential energy landscape whose symmetries and
features can be programmed precisely.
Static optical landscapes are useful for templating
the crystallization of uniformly sized colloidal particles, and more generally in
modifying such dispersions' phase transitions and dynamics
\cite{korda02a,mangold03}.
How individual colloidal particles navigate such landscapes when driven by an
external force has
proved a surprisingly challenging problem with immediate technological applications.

\begin{figure}[htbp]
  \centering
  \includegraphics[width=3in]{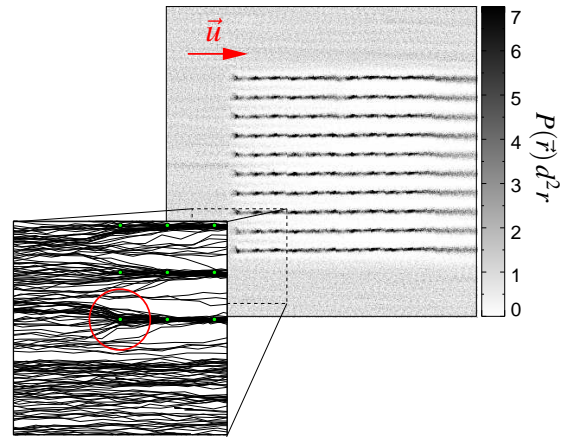}
  \caption{Measured trajectories of fluid-borne micrometer-diameter
    colloidal particles
    encountering a $10 \times 10$ square array of holographic optical
  traps with a lattice constant of $3~\micron$n.}
  \label{fig:h0}
\end{figure}

Figure~\ref{fig:h0} shows the measured \cite{crocker96,korda02b} trajectories of colloidal spheres
as they are carried by flowing water through a $10 \times 10$ square array of holographic optical traps.
At flow speeds of $u = 50~\unit{\micron/sec}$, the force due to viscous drag on the $\sigma = 1.5~\micron$ diameter
silica spheres is comparable to the individual tweezers' maximum trapping force.  Although particles are drawn
toward the rows of tweezers from a range comparable to their diameter, they hop freely from trap to trap 
along the array's $[10]$ axis.
The particles' channeling along the rows is clearly demonstrated by the compiled
probability density $P(\vec{r}) \, d\vec{r}$ for finding particles within $d\vec{r}$ of $\vec{r}$ relative
to the bulk, which also is plotted in Fig.~\ref{fig:h0}.

Tilting the array so that its $[10]$ axis no longer is aligned with the flow
presents the particles with an opportunity to escape from their commensurate paths through the
potential energy landscape.  Over some range of angles, however, a particle can be deflected enough
by its encounter with one trap to fall into the domain of attraction of the next.  In this case,
the particle's trajectory can remain kinetically locked in to the commensurate direction through the
landscape \cite{korda02b}.

\begin{figure}[thbp]
  \centering
  \includegraphics[width=2.5in]{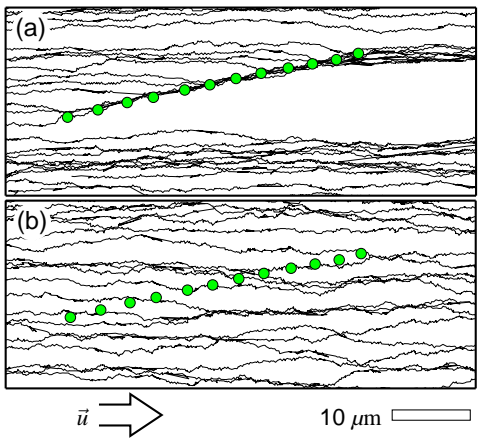}
  \caption{Optical fractionation.  (a) Measured trajectories of large colloidal
  spheres, 1.58~\micron in diameter, dispersed in water flowing at 
  50~\unit{\micron/sec}, are deflected by a line of 12
  holographic optical traps, whose positions are indicated by circles.
  (b) Smaller 1~micron diameter spheres dispersed in the same flow are
  not deflected (b).}
  \label{fig:fractionation}
\end{figure}

Fluid-borne objects can become kinetically locked-in
to even a single inclined line of traps
\cite{ladavac04,pelton04a}, as the data in
Fig.~\ref{fig:fractionation} show.
This effect has immediate technological
implications.
Because the potential energy landscape experienced by a passing
particle depends sensitively on the object's optical form factor
\cite{pelton04a}, small differences in size, shape or composition
can cause different objects to follow radically different
paths through an optical trap array.
The resulting spatial separation
is the basis for a continuous
and continuously tunable sorting process known as optical
fractionation 
\cite{macdonald03,ladavac04}.

Although the theory for transport through optical tweezer arrays
is not yet complete \cite{pelton04a,gopinathan04}, preliminary results
suggest
that optical fractionation may be able to sort flowing objects
with exponential size selectivity \cite{ladavac04,pelton04a}.
Two-dimensional and three-dimensional arrays, moreover, exhibit
a rich hierarchy of kinetically locked-in commensurate pathways
\cite{korda02b,gopinathan04}, all of which can be exploited for
multi-channel sorting.

The first generation of experiments \cite{korda02b,macdonald03,ladavac04} 
has relied
on driving forces exerted by flowing fluid.
Symmetry-guided transport also should arise for particles driven
by electrophoresis, electroosmosis and related mechanisms.
Field-driven optical fractionation will make possible sorting
on the basis of such properties as surface charge density and should
be useful for processes monitoring and control and could provide
the basis for a new family of analytical chromatographies.

\section{Active landscapes: conveyors and ratchets}

While static optical trap arrays act as filters or prisms
for externally driven dispersions, dynamic arrays are useful for inducing motion.
Dynamic holographic optical traps do not move continuously, as do 
optical tweezers scanned with moving mirrors \cite{sasaki91} or traps
created by the generalized phase contrast method \cite{mogensen00}.
Rather, one pattern of traps dissolves into another as the DOE encoding
the first is replaced by that encoding the second.
If trapped objects diffuse slowly enough, they still can be
passed from trap to trap by rapidly updating the phase hologram.
Viscous relaxation, in this case, plays the role in active holographic
transport that persistence of vision plays in cartoon animation to provide
the appearance of continuous motion.

Sequences of overlapping trapping patterns can dynamically organize mesoscopic objects
into arbitrary three-dimensional configurations, and reorganize them quasi-continuously
\cite{reicherter99,liesener00,curtis02,sinclair04,jordan04,leach04,roichman05}.
Periodically cycled sequences of as few as three holograms can induce complicated
patterns of motion over large areas through a process called
optical peristalsis \cite{koss03}.   Here, an object is transferred forward
from one manifold of traps in a given pattern to the next by
two or more intervening trapping patterns whose manifolds bridge the gap.
The sequence of patterns breaks spatiotemporal symmetry and ensures that
motion proceeds in the intended direction.
Unlike interactive manipulation that requires an individual particle to
be captured and its path to be calculated, optical peristalsis operates
over the entire field of view, directing and orienting objects automatically
through small sequences of precalculated holograms, much like an optical
conveyor belt.

This process also provides a means to implement a so-called thermal ratchet
\cite{reimann02}, in which diffusing particles' random Brownian motion is
rectified into a directed flux by a time-evolving potential energy landscape.
Unlike conventional motors and deterministic processes such as optical peristalsis
whose performance is degraded by random
fluctuations, thermal ratchets are stochastic machines and \emph{require} noise to operate.
Most proposed models for thermal ratchets exploit a space-filling
spatially asymmetric potential energy landscape.
Breaking spatial symmetry is not enough to eke a flux out of fluctuations for a system
in equilibrium.
As part of a sequence of states driving the system out of equilibrium, however,
it can help to break diffusion's spatiotemporal symmetry and thereby induce motion.
This works even if the landscape itself has no
overall slope and thus exerts no net force.

\begin{figure}[htbp]
  \centering
  \includegraphics[width=4in]{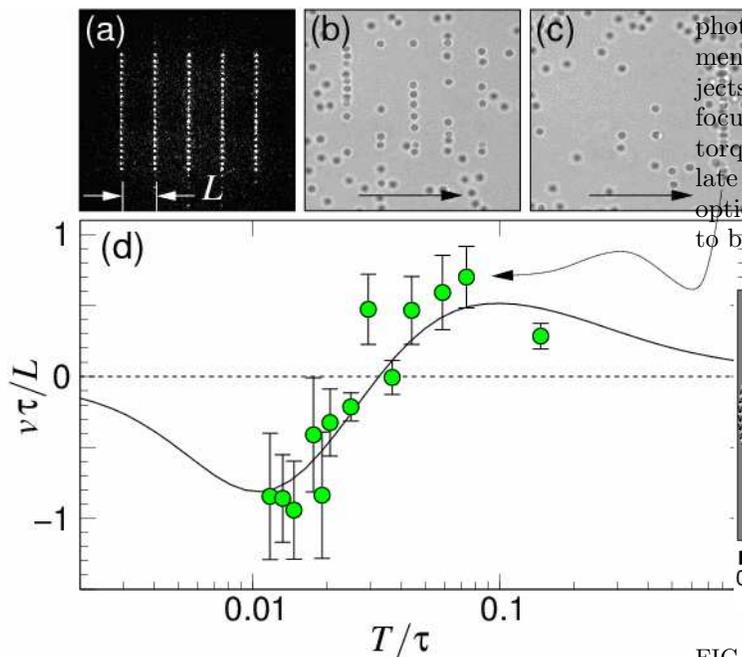}
  \caption{A thermal ratchet implemented with holographic optical
  tweezers. (a) The focused light from a $20 \times 5$ array with
  manifolds separated by $L = 3.8~\micron$.  (b) A dispersion of 1.58~\micron
  diameter spheres interacting with the array. (c) After repeated
  displacements of the array by $L/3$ and $2L/3$, with each step
  lasting $T = 5~\unit{sec}$, all the spheres are translated to the
  right.
  (d) The transport velocity $v$ as a function of dwell time shows
  flux reversal as the cycle rate increases.}
  \label{fig:ratchet}
\end{figure}
A regular array of holographic optical tweezers, such as that shown in
Fig.~\ref{fig:ratchet}, presents a potential energy landscape
that neither fills space nor breaks spatial symmetry.
Even the individual traps are locally symmetric potential wells.
Nevertheless, translating the array first by one third of a lattice constant, then
by two, and then returning it to its initial state creates a
discrete-state traveling ratchet \cite{lee05} whose time evolution
breaks spatiotemporal symmetry.  The resulting motion differs
from that induced by optical peristalsis in a way that leads
to additional applications.

If the lattice constant $L$ is comparable to the traps'
effective widths \cite{pelton04a}, then the traveling ratchet 
reduces to an example optical peristalsis,
and particles are deterministically translated along the displacement
direction.  Increasing the separation causes particles trapped
in one state to be left behind in a flat and featureless region of
the potential energy landscape in the next state.  They must diffuse to the
nearest
manifold of traps before they can be localized and transported.
If the time $\tau$ required for the particles to diffuse across the
potential energy plateau is shorter than the duration $T$ of each
state, then most particles are transported forward.  On the other hand,
if the particles
diffuse too slowly, they can miss the
forward-going wave and may end up instead being transported
\emph{backward} \cite{lee05}.  
Such flux reversal as a function of cycle time $T$
and trap separation $L$ is a hallmark of thermal ratchet operation,
and is clearly seen in the data in Fig.~\ref{fig:ratchet}.

Flux reversal in microfabricated thermal ratchets already has been
exploited for separating DNA and other macromolecules on the basis
of their diffusivity \cite{hughes02}.  The holographically implemented
variant complements optical fractionation by permitting automatic
sorting \emph{in situ}.  Variants of the holographic optical ratchet
exploit more subtle symmetries to achieve simpler operation \cite{lee05a}
or more sophisticated configurations of tweezers to optimize sorting.

\section{Multimode traps}
\label{sec:multimode}

The previous sections addressed some of the applications for
holographic arrays of conventional optical tweezers.  Wavefront
engineering through Eq.~(\ref{eq:transfermatrix}) provides access to
more sophisticated traps.  For example, the deceptively
simple phase profile, $\varphi_\ell(\vecr) = \ell \theta$, where
$\theta$ is the azimuthal angle around the optical axis and $\ell$ is
an integer winding number, transforms a conventional TEM$_{00}$ mode
into a helical mode \cite{allen92}.  The axial screw dislocation introduced by this
phase profile leads to perfect destructive interference along the beam's axis.
A helical beam consequently focuses to a \emph{dark} spot, its intensity
being redistributed into an annulus whose radius \cite{sacks98,curtis03,sundbeck05}
scales with the topological charge $\ell$.  Each photon in a helical model
carries $\ell \hbar$ orbital angular momentum \cite{allen92,soskin97} that it can
transfer to illuminated objects \cite{he95a}.
The ring-like optical trap that results from focusing a helical mode
\cite{he95,gahagan96,simpson96} therefore can exert torques on trapped
objects \cite{he95a}, causing them to circulate around the ring \cite{oneil00,oneil02,curtis03}.
Such torque-exerting optical traps, and their generalizations \cite{curtis03a}, have
come to be known as optical vortices.

\begin{figure}[htbp]
  \centering
  \includegraphics[width=3in]{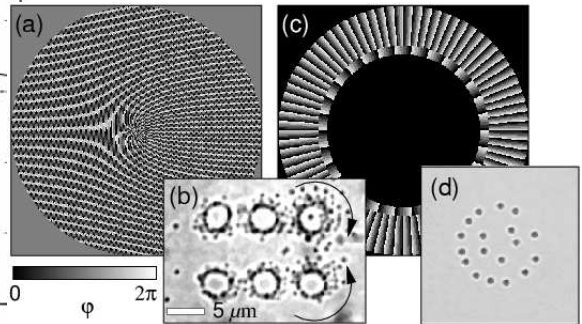}
  \caption{(a) Phase mask encoding a $3 \times 2$ array of
  counter-rotating optical
  vortices with $\ell = \pm 30$.  
  (b) Microoptomechanical pump created by projecting the vortex array
  into a colloidal dispersion.  Fluid flows from right to left as the
  rings of particles circulate.  (c) Phase mask encoding concentric
  optimized optical vortices.  (d) A microoptomechanical Couette
  rheometer 
  created by projecting the concentric vortices into a colloidal dispersion.}
  \label{fig:vortex}
\end{figure}
The vortex-forming phase profile can be imposed on an individual trap in an array
in much the same way as the displacement-inducing curvature described by Eq.~(\ref{eq:displace}).
As a result, a single DOE such as the example in
Fig.~\ref{fig:vortex}(a)
can create arrays of optical vortices, each
with independently specified positions, intensities, and topological
charges.
Figure \ref{fig:vortex}(b) shows 
the array of $\ell = +30$ and $\ell = -30$ optical vortices encoded by
this DOE trapping and 700~\unit{nm} diameter silica
spheres.
The rapidly circulating spheres entrain flows in the suspending water,
which
results in a steady stream along the axis of the array \cite{ladavac04a}.

Unlike conventional trap-forming holograms, the helical phase profile 
$\varphi_\ell(\vec{r})$ has a topological defect at the origin.
Regions of the DOE nearest the singularity contribute to an
optical vortex's intensity at its outer edge, while more removed
regions contribute to its inner edge \cite{guo04}.  
This bright inner edge, moreover,
is where an optical vortex traps and circulates particles.
Removing the central region of a helical mode-former, as in
Fig.~\ref{fig:vortex}(c) does not affect the optical vortex's
performance as a torque-exerting optical trap, but reduces the
amount of wasted light.  The empty central region then can play host
to one or more additional optical vortices \cite{guo04,ladavac05}, as
shown in Fig.~\ref{fig:vortex}(d).  These concentric rings act as
a microscopic Couette rheometer useful for studying viscoelastic
properties at the nanometer to micrometer scale \cite{ladavac05}.

More general superpositions of helical modes give rise to other
microoptomechanical devices, such as optical cogwheels
\cite{jesacher04} that are useful for sorting objects by size,
modulated optical vortices \cite{curtis03a} that project objects
on complicated trajectories through the focal plane.

A conical phase profile transforms an optical tweezer into a
diffractionless Bessel beam \cite{arlt01,garceschavez02} that
can stack objects into three dimensional columns.  Adding
helical and conical profiles creates generalized Bessel-Laguerre traps
that also exert both torques and forces over extraordinarily larger
ranges.  Combining them by multiplication generates spiral intensity
patterns whose trapping applications have yet to be full investigated
\cite{alonzo05}.

It should be emphasized that all of the microscopic manipulations
described in this article, and a great many more, can be accomplished
with a single optical train.  Strikingly different functionalities 
result from changes in the transfer matrices in
Eq.~(\ref{eq:transfer}), all under software control.  Some already
have provided new avenues for fundamental research.  Others are
making inroads into industrial processes.  In the broadest
sense, holographic optical trapping provides a powerful and very
general approach to interacting with the microscopic world.  Progress
in both the technique and its real-world applications should be rapid.

This work was supported by NSF Grants number
DBI-0233971 and DMR-0451589.


\end{document}